\pdfoutput=1
\documentclass[sigconf]{acmart}

\AtBeginDocument{%
  }

\AtBeginDocument{%
  }

\setcopyright{none}

\renewcommand\footnotetextcopyrightpermission[1]{}
\usepackage{tikz}
\usetikzlibrary{positioning, fit, arrows.meta, decorations.pathreplacing, backgrounds}
\usepackage{enumitem}
\usepackage{xcolor}
\usepackage{booktabs}
\usepackage{siunitx}
\usepackage{lipsum} 
\usepackage{balance}

\newcommand{\pkg}{JaxMARL-BC}
\begin{document}

\title{JaxMARL-BC: Multi-Agent Reinforcement Learning \\ for Macroeconomic Modelling, in Seconds}

\author{Daniele Montagnani}
\affiliation{
  \institution{Università di Pavia}
  \country{Italy}
}
\affiliation{
  \institution{USI Lugano}
  \country{Switzerland}
}
\orcid{0009-0004-9957-2320}

\author{Andrea Coletta}
\affiliation{
  \institution{Banca d'Italia\textsuperscript{\textdagger}}
  \country{Italy}
}
\orcid{0000-0003-1401-1715}

\author{Antonietta Mira}
\affiliation{
  \institution{USI Lugano}
  \country{Switzerland}
}
\affiliation{
  \institution{Università dell'Insubria}
  \country{Italy}
}
\orcid{0000-0002-5609-7935}

\author{Aldo Glielmo}
\authornote{\href{mailto:aldo.glielmo@bancaditalia.it}{aldo.glielmo@bancaditalia.it}}
\affiliation{
  \institution{Banca d'Italia\textsuperscript{\textdagger}}
  \country{Italy}
}
\orcid{0000-0002-4737-2878}

\renewcommand{\shortauthors}{Montagnani, Coletta, Mira, Glielmo}

\begin{abstract}
Multi-agent reinforcement learning (MARL) offers a principled way of populating simulated economies with many heterogeneous decision makers. 
Its adoption, pioneered in finance, has nonetheless been slow in macroeconomics because of two main limitations: the lack of a shared frame of reference connecting MARL systems to dynamic stochastic general equilibrium (DSGE) models the profession already trusts, and the very long training times.
The first limitation was recently addressed by the Multi-Agent Reinforcement Learning Business Cycle (MARL-BC), which places deep RL households in a real business cycle economy and reduces to classic DSGE models in specific limits.
Here we address the second limitation.
We present \pkg{}, a JAX-accelerated simulator that trains MARL-BC economies in seconds (between $75\times$ and $724\times$ faster than the reference implementation), and scales to $200{,}000$ independent learning households. 
We verify its correctness against the known DSGE limits, characterise its scaling through a simple cost law, and demonstrate its value by a proof-of-concept calibration of an extended model with heterogeneous returns on capital, in a two-hour experiment that would previously have taken over a month. 
We release \pkg{} as open source.
\end{abstract}



% \ccsdesc[500]{Applied computing~Economics}
% \ccsdesc[500]{Computing methodologies~Modeling and simulation~Simulation types and techniques~Agent / discrete models}
% \ccsdesc[500]{Artificial intelligence~Planning and scheduling~Planning under uncertainty}
% \ccsdesc[500]{Machine learning~Machine learning approaches~Markov decision processes}


\maketitle
\pagestyle{plain}

\begingroup
\renewcommand{\thefootnote}{\textdagger}
\footnotetext{The views expressed in this paper are those of the authors and do not necessarily reflect those of Banca d'Italia.}
\let\thefootnote\relax
\footnotetext{This article was published in the \emph{Proceedings of the 7th ACM International Conference on AI in Finance (ICAIF '26)}, Milan, Italy, and is also available at \url{https://doi.org/10.1145/3858063.3858703}.}
\endgroup

\section{Introduction}
% RL AND MARL AS A NEW PARADIGM
Reinforcement learning (RL)~\cite{sutton1998reinforcement} and multi-agent RL (MARL)~\cite{albrecht2024multi} have gradually become the basis of a new modelling paradigm for simulating the behaviour of agents in economics and finance.
In RL, agents learn to maximise their rewards through repeated interaction with the economy they inhabit, so that MARL becomes a principled way of populating a simulated economy with many diverse decision makers, each optimising a reasonable objective.

% ADOPTION QUICK IN FINANCE SLOW IN MACRO
The adoption of this approach was pioneered in finance, where RL and MARL are now commonly used to represent trading, hedging or market-making agents across a range of markets~\cite{karpe2020multi,ardon2021towards,gao2023deeper,mascioli2024financial,Evans2024,bacaloni2025natural,wang2025multi}, or within multi-agent market simulators supported by a growing ecosystem of open-source platforms~\cite{amrouni2021abides,liu2021finrl,Phantom2023,frey2023}.
In macroeconomics and structural macro-financial modelling, adoption has been comparatively slower.
Indeed, in spite of a growing body of specialised research in the field, from stylised economies~\cite{zheng2022ai,cook2025reinforcement,brusatin2024simulating}, to proofs of concept in monetary policy~\cite{chen2021deep,hinterlang2021optimal}, and tax design~\cite{CurryEtAl2023,MiEtAl2023,dwarakanath2024tax,agrawal2025robust}, the methodology has not entered the toolkit of policymakers and practitioners.

% TWO REASONS FOR SLOW ADOPTIONS
In our view, the reason is twofold.
On the one hand, this is due to scepticism and resistance towards alternative modelling methodologies, from a community that is long accustomed to a specific way of building models based on the dynamic stochastic general equilibrium (DSGE) approach.
This resistance is understandable, as the DSGE approach provides a shared frame of reference that keeps modelling out of the ``wilderness of bounded rationality''~\cite{farmer2024making},
where infinitely many deviations from optimality become available.
On the other hand, this is due to the considerably higher computational cost of training RL or MARL systems, whereas a typical DSGE model can be specified and solved in seconds.
A shared framework and fast solvers, in turn, make these models practical to use, and explain the remarkable success of specialised modelling software such as Dynare~\cite{adjemian2011dynare,villemot2011solving}.
% %
% Ultimately, making models practical to use, and they explain the remarkable success of specialised modelling software such as Dynare~\cite{adjemian2011dynare,villemot2011solving}.
%

% FIRST LIMITATION RECENTLY ADDRESSED, SECOND STILL THERE
The first limitation of MARL approaches has been convincingly addressed in a recently introduced framework, the Multi-Agent Reinforcement Learning Business Cycle (MARL-BC)~\cite{gabriele2026}.
MARL-BC places deep RL households inside a real business cycle (RBC) economy, so that the MARL system reduces to classic DSGE models in specific limits, recovering the representative-agent RBC and the mean-field Krusell--Smith economy.
Hence, MARL-BC extends the classic models rather than replacing them, and it does so in a way that can be checked against results the profession already trusts.
However, MARL-BC does not address the second limitation we mentioned, since training requires multiple hours per run~\cite{gabriele2026}.

% OUR CONTRIBUTION ADDRESSES THE SECOND LIMITATION
In this work, we leverage JAX~\cite{jax2018}, a library for just-in-time compilation of numerical code to hardware accelerators, to reduce MARL-BC training time from hours to seconds, thereby addressing the second adoption bottleneck described above, and fostering a wider usage of the methodology by economists and practitioners.
JAX has often been proposed as a viable way to lower RL training times by allowing environment, policy and optimiser to live inside a single compiled program directly on the accelerator~\cite{lu2022discovered,flair2024jaxmarl,pretorius2021mava}.
The financial community has already extensively explored this route, with the JAX-LOB simulator for limit order book RL trading agents~\cite{frey2023}, and the JaxMARL-HFT simulator extending it to large-scale multi-agent high-frequency trading~\cite{mohl2025jaxmarl}.
EconoJax~\cite{ponse2024} applies the same strategy to a stylised economic simulation, but no equivalent exists for structural MARL-driven macroeconomic modelling, a gap this paper fills.

% LIST OF CONTRIBUTIONS
More specifically, our contributions are as follows:
\begin{itemize}[leftmargin=15pt, itemsep=2pt, topsep=2pt, parsep=0pt]
    \item We propose \pkg{}, an end-to-end JAX simulator which trains MARL-BC economies in seconds rather than hours, and scales easily to hundreds of thousands of independent learning agents.
    \item We verify the correctness of our simulator against the known DSGE limits of the MARL-BC, reproducing the results at ten times the population size and at a small fraction of the cost.
    \item We extensively measure the scalability to large populations of agents and to large numbers of parallel environments, which we characterise via a simple cost law that we verify empirically.
    \item We showcase the capability of the simulator by performing a realistic calibration exercise of a large model with heterogeneous returns on capital, with $110$ training runs sweeping different degrees of ex-ante heterogeneity in a two-hour computation that would have previously taken over a month.
\end{itemize}
%

% AIM TO IMPROVE ADOPTION, HENCE OPEN SOURCE AND DASHBOARD
We release \pkg{} as open source\footnote{\url{https://github.com/danmonuni/jax-marl-bc}.}, together with a prototype LLM-augmented dashboard that turns natural-language requests into simulation configurations, so that non-technical users can experiment without touching the command line.
Removing the computational barrier to MARL in macroeconomics can let the methodology spread beyond toy demonstrations and become common practice for structural macro and macro-financial modelling.
 
\vspace{0.2cm}
We structure the rest of this work as follows.
In Sec.~\ref{sec:related_work} we review related work.
In Sec.~\ref{sec:simulator} we describe the \pkg{} simulator.
%
% In Sec.~\ref{subsec:model} and Sec.~\ref{subsec:speedup} we present the model underlying the simulator and the way in which we achieve the JAX speedup, respectively.
%
In Sec.~\ref{sec:setup} and Sec.~\ref{sec:results} we report the experimental setup and the results of our experiments.
Finally, in Sec.~\ref{sec:conclusion} we conclude.

\section{Related work}
\label{sec:related_work}

Training RL and MARL systems typically requires substantial time and computational resources, which can limit their adoption. 
The creation of simulation environments in JAX has therefore become an active line of work.
The design pattern was popularised for single-agent RL by PureJaxRL~\cite{lu2022discovered} and is now packaged in multi-agent libraries such as JaxMARL~\cite{flair2024jaxmarl} and Mava~\cite{pretorius2021mava}, which report speedups of one to four orders of magnitude over comparable CPU runs.
Among application domains, finance has been the most active adopter of this pattern.
JAX-LOB~\cite{frey2023} was the first GPU-accelerated limit order book simulator, processing thousands of books in parallel and delivering at least a $7\times$ speedup over comparable CPU implementations.
JaxMARL-HFT~\cite{mohl2025jaxmarl} builds on JAX-LOB and extends it to a heterogeneous set of agents with distinct observations, actions and rewards, reporting up to a $240\times$ reduction in training time.
EconoJax~\cite{ponse2024} ports the AI-Economist environment~\cite{zheng2022ai} to JAX and trains populations of $100$ agents in fifteen minutes, down from the several days required by the original implementation.
\pkg{}  follows the same design philosophy, but targets a
domain none of these packages cover, namely structural heterogeneous macroeconomic modelling.
Our work is also close to the ABIDES-Economist~\cite{dwarakanath2024abides}, which populates agent-based economies with RL agents, but which is not implemented in JAX and does not specifically target established macroeconomic benchmarks.

% NON-RL SOLVERS FOR HETEROGENEOUS-AGENT MODELS
A different line of work uses deep learning, without RL, to solve heterogeneous-agent models~\cite{maliar2021deep,azinovic2022deep,han2026deepham}, and certain fast conventional methods, such as the sequence-space Jacobian~\cite{auclert2021using}, can also be used for the same task.
Closer to our approach, structural reinforcement learning~\cite{yang2025structural} has recently brought RL into heterogeneous-agent models, proposing a solution method for mean-field games that exploits structural RL simulations.
\pkg{} does not aim to compete with these methods on the classes of models they already solve well. 
Rather, it aims to facilitate the growth and adoption of a new modelling paradigm based on independent learning agents, which these methods do not represent.

\begin{figure*}[t]
    \centering
    \resizebox{0.65\columnwidth}{!}{%
    \begin{tikzpicture}[
        font=\small,
        >={Stealth},
        lbl/.style={font=\small\itshape},
    ]
    \def\W{2.3}
    \def\H{2.3}
    \def\dx{0.22}
    \def\dy{0.15}
    \foreach \i in {3,2,1} {
      \begin{scope}[shift={(\i*\dx,\i*\dy)}]
        \draw[gray!50, thin] (0,0) rectangle (\W,\H);
      \end{scope}
    }
    \fill[blue!8] (0,0) rectangle (\W,\H);
    \draw[blue!45!black, thick] (0,0) rectangle (\W,\H);
    \foreach \gx in {1,...,6} { \draw[blue!25] ({\gx*\W/7},0) -- ({\gx*\W/7},\H); }
    \foreach \gy in {1,...,3} { \draw[blue!25] (0,{\gy*\H/4}) -- (\W,{\gy*\H/4}); }
    \node[blue!45!black, font=\tiny\itshape, anchor=north west] at (0.06,\H-0.06) {$t{=}0$};
    \node[gray!65, font=\tiny\itshape, anchor=south west] at ({3*\dx},{\H+3*\dy+0.01}) {$t{=}T$};
    \draw[decorate, decoration={brace, amplitude=5pt, mirror}] (0,-0.12) -- (\W,-0.12);
    \node[align=center, lbl, green!30!black, font=\footnotesize] at (\W/2,-0.75)
      {$n$ agents\\ array axis};
    \draw[decorate, decoration={brace, amplitude=5pt, mirror}] (-0.12,\H) -- (-0.12,0);
    \node[align=center, lbl, blue!45!black, font=\footnotesize, rotate=90] at (-0.95,\H/2)
      {$E$ environments\\ \texttt{jax.vmap}};
    \draw[->, thick, red!55!black] (\W,\H) -- ({\W+3*\dx},{\H+3*\dy});
    \node[align=center, lbl, red!55!black, font=\footnotesize, anchor=west] at ({\W+3*\dx-0.05},{\H+3*\dy-0.55})
      {$T$ steps\\ \texttt{lax.scan}\\ sequential};
    \node[align=center, font=\small\bfseries] at ({\W/2+3*\dx/2},{\H+3*\dy+0.55})
      {$W = E \times n$};
    \begin{scope}[on background layer]
    \draw[rounded corners=4pt, dashed, thick, gray!55]
      (-1.505,-1.05) rectangle ({\W+2.285},{\H+3*\dy+0.9});
    \end{scope}
    \node[anchor=north west, font=\tiny\ttfamily, gray!45!black] at (-1.455,{\H+3*\dy+0.85}) {jit(train)};
    \end{tikzpicture}%
    }
    \hspace{0.3cm}
    \includegraphics[width=0.6\columnwidth]{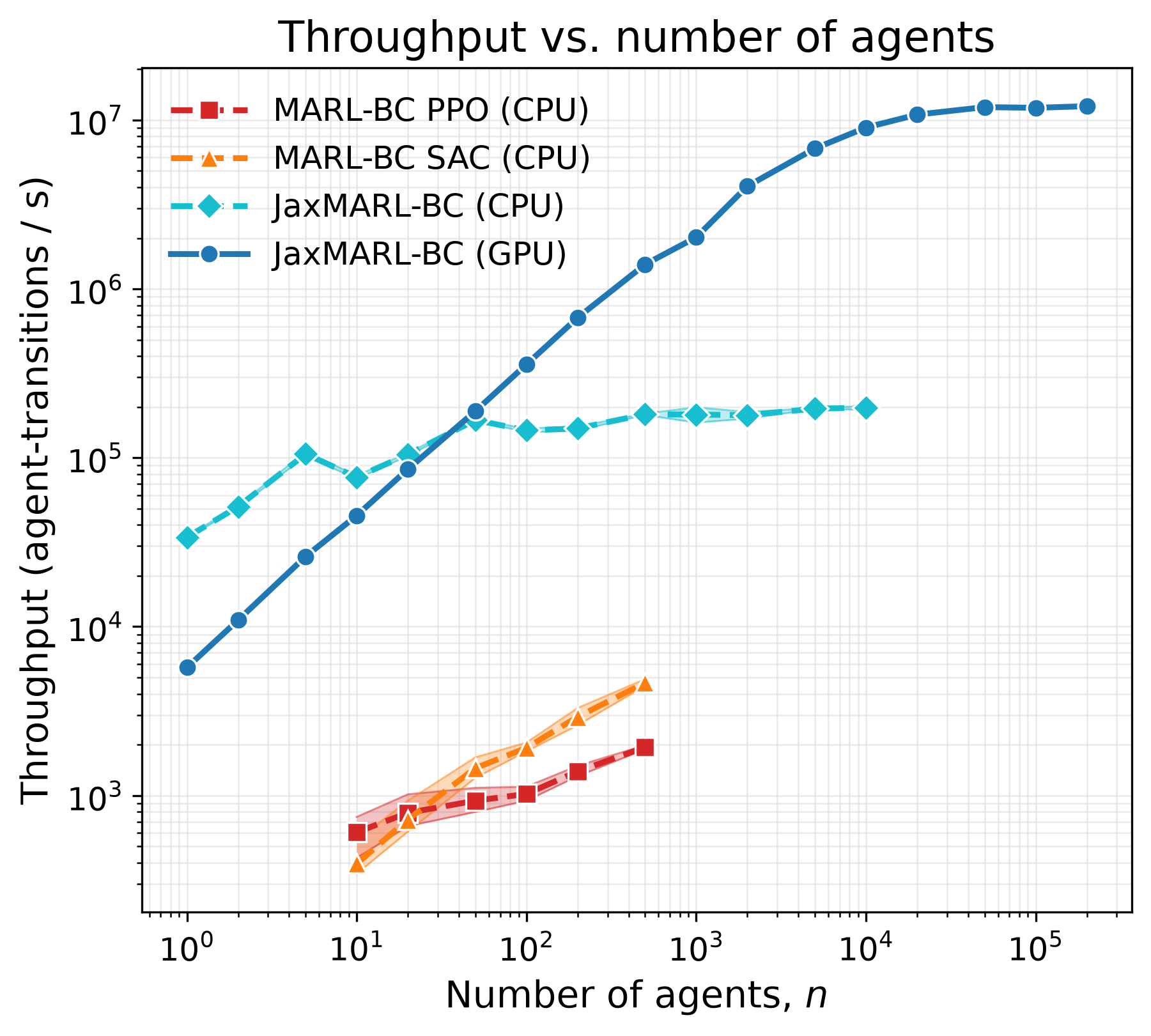}
    \hspace{0.3cm}
    \includegraphics[width=0.6\columnwidth]{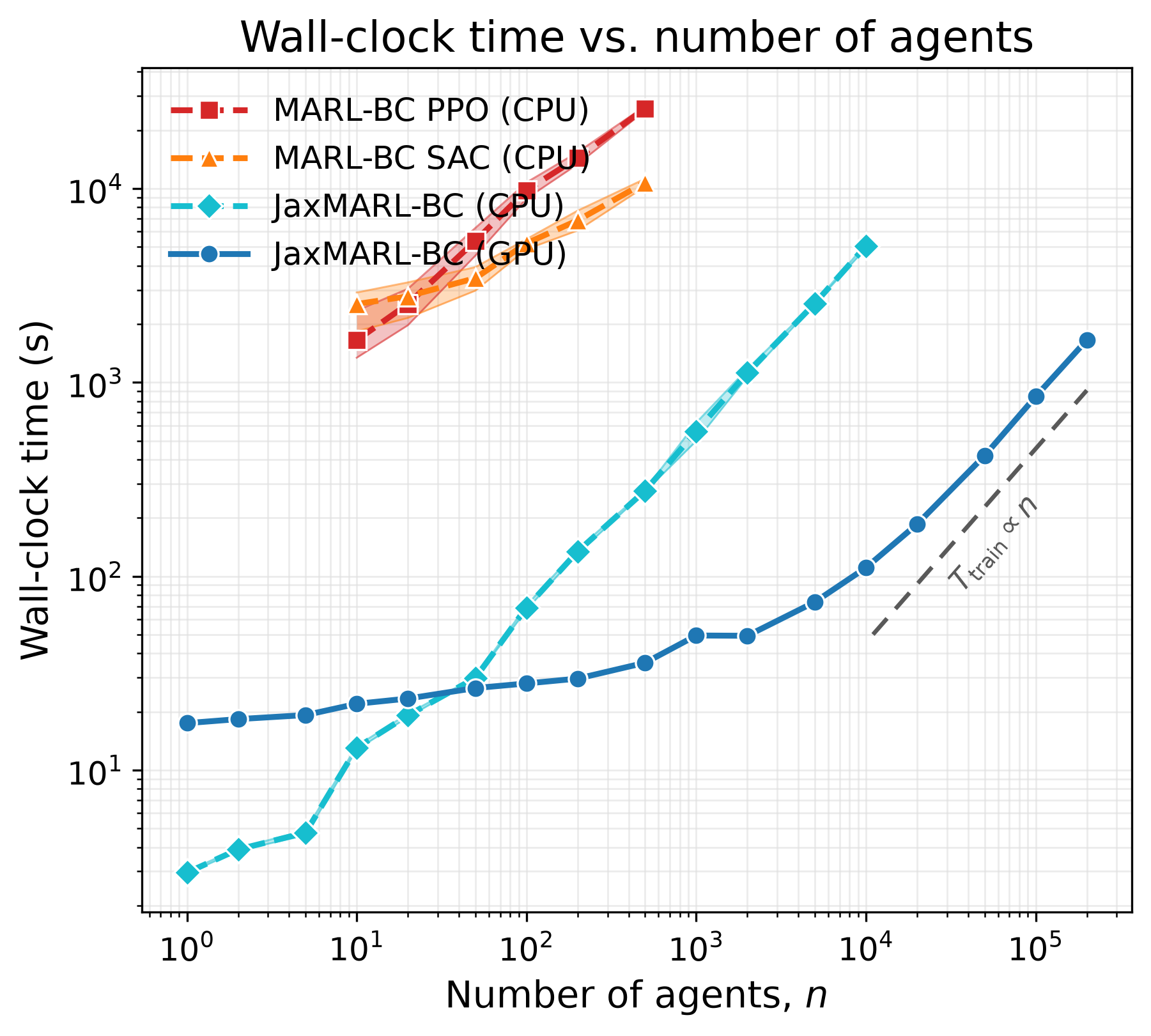}
    \caption{Speedup illustration and resulting scaling with population size $n$.
    \normalfont
    \emph{Left}: an illustration of the axes of parallelism.
    One rollout evaluation is a rectangular array of agent-transitions with $n$ agents along one axis (a plain array dimension) and $E$ environments along the other (with \texttt{jax.vmap}).
    This is advanced sequentially over $T$ rollout periods, within a \texttt{lax.scan}, which is not parallelisable.
    Rollouts and training steps live inside one \texttt{jit}-compiled program.
    \emph{Centre and right}: throughput and wall-clock time of training runs with $10^5$ periods each, as a function of the number of agents $n$.
    Curves are shown for both CPU and GPU backends (dashed light blue and solid blue lines) and each plots the mean and standard deviation over three seeds.
    The right-most panel also reports a linear slope as a dashed grey line.
    The simulator is evaluated against comparable MARL-BC runs (orange and red lines) with times taken from the original paper.
    }
    \label{fig:scaling-vs-nagents}
\end{figure*}

\section{The \pkg{} simulator}
\label{sec:simulator}

We describe \pkg{} in two parts.
In the first part (Sec.~\ref{subsec:model}) we provide an overview of the MARL-BC model that forms the basis of the simulator, and in the second part (Sec.~\ref{subsec:speedup}) we describe the architectural choices that allow for drastically lower training time, and we outline a simple cost model for training time.
Finally, in Sec.~\ref{subsec:dashboard} we briefly describe the LLM-augmented dashboard which allows non-technical users to experiment with the simulator.

\subsection{The underlying model}
\label{subsec:model}

The underlying model is based on the MARL-BC economy of~\cite{gabriele2026}.

\vspace{0.2cm}
\noindent
\textbf{One period.}
Household $i$ enters period $t$ with capital $k^i_t$, a labour endowment $\ell^i_t$, and fixed productivities of capital $\kappa^i$ and of labour $\lambda^i$.
Effective aggregates are productivity-weighted population means,
\begin{equation}\label{eq:agg}
K_t = \frac{1}{n}\sum_{i=1}^{n} \kappa^i k^i_t,
\qquad
L_t = \frac{1}{n}\sum_{i=1}^{n} \lambda^i \ell^i_t,
\end{equation}
which are turned into output through a Cobb--Douglas function
\begin{equation}\label{eq:production}
Y_t = A_t K_t^{\alpha} L_t^{1-\alpha},
\end{equation}
with capital elasticity $\alpha$ and aggregate technology $A_t$.
The returns on capital and wages of the households depend on their productivity of capital and labour via
\begin{equation}\label{eq:prices}
r^i_t = \alpha \frac{Y_t}{K_t}\kappa^i,
\qquad
w^i_t = (1-\alpha) \frac{Y_t}{L_t}\lambda^i,
\end{equation}
and the cash-on-hand at time $t$ is given by
\begin{equation}\label{eq:cash}
a^i_t = w^i_t \ell^i_t + r^i_t k^i_t + (1-\delta) k^i_t,
\end{equation}
where $\delta$ is the capital depreciation rate.

\vspace{0.2cm}
\noindent
\textbf{The household's problem.}
The households choose their consumption fraction $\hat{c}^i_t$, from which consumption and next-period capital follow as
\begin{equation}\label{eq:split}
c^i_t = \hat{c}^i_t\, a^i_t,
\qquad
k^i_{t+1} = (1-\hat{c}^i_t)\, a^i_t.
\end{equation}
The households can also choose next-period labour endowment $\ell^i_{t+1}$.
Consumption and labour enter the reward as
\begin{equation}\label{eq:reward}
R^i_t = \log c^i_t + b \log\!\left(1 - \ell^i_t\right),
\end{equation}
where the weight $b \ge 0$ prices leisure relative to consumption.
Each household thus learns a policy that maps its own observations to an action $(\hat{c}^i_t, \ell^i_{t+1})$ in order to maximise $\mathbb{E}[\sum_t \beta^t R^i_t]$, where $\beta$ is the household's discount factor.

\vspace{0.2cm}
\noindent
\textbf{Learning.}
Households are trained by deep RL, and specifically by Proximal Policy Optimization (PPO)~\cite{schulman2017proximal}.
PPO is a standard algorithm for RL, and also for MARL where typically, as in our case, all agents are treated as independent learners in the same environment~\cite{de2020independent}.
With $n>1$ learners in one economy, the transitions and rewards any household observes depend on what the others are currently doing, so each faces a non-stationary problem and the single-agent convergence guarantees no longer apply~\cite{albrecht2024multi}.
For symmetric agents, a standard remedy is parameter sharing, in which a single policy network serves the entire population and is evaluated on each agent's private observations.
This stabilises training without forcing identical behaviour, since household characteristics enter through the observation~\cite{yu2022}.
MARL-BC interprets this configuration as a stylised form of ``social learning''~\cite{whalen2018sensitivity,ndousse2021emergent}, and we adopt the same approach here.

\vspace{0.2cm}
\noindent
\textbf{Known limits of the model.}
The MARL-BC paper shows that the model reproduces standard DSGE models in two limit cases. 
Specifically, with a single agent ($n=1$), individual and aggregate quantities coincide and the framework reduces to the textbook representative-agent RBC, while with many ex-ante identical agents ($n\gg1$ and $\kappa^i = \lambda^i =1 \; \forall i$), the framework reproduces the well-known mean-field Krusell--Smith model. 
These two limits make the present work easily testable, since the target behaviour is known from a well-established literature, and they also address the first of the adoption obstacles discussed in the introduction, by anchoring the framework to model classes widely recognised by the community.

\subsection{The JAX speedup}
\label{subsec:speedup}
We now describe the JAX design that enables the speedup of our simulator.
The left panel of Figure~\ref{fig:scaling-vs-nagents} shows the three axes upon which a single rollout of the simulator is parallelised and executed.
These axes differ in the kind of dependence that holds among their elements, and each kind calls for a different JAX construct.
Along the first axis are the $n$ households of one economy, which are \emph{not} independent of one another, as they are coupled through the aggregates of Eq.~\eqref{eq:agg} that determine the prices they all face.
The agent axis is therefore a reduction axis, and we represent it as a plain array dimension so that $n = 10$ and $n = 10{,}000$ execute the very same program, only on wider arrays.
By contrast, the second axis represents $E$ copies of the economy, which are fully independent of one another.
This is the kind of parallelism \texttt{jax.vmap} is designed for, as it takes a function written for a
single instance and lifts it to a batch of independent ones.
Here, a single \texttt{jax.vmap} over the environment's `reset' and `step' functions produces the $E$ parallel economies.
The policy forward pass flattens these two axes into one batch dimension, so the policy network only ever sees their product, the batch width $W = E \times n$.
The third axis, the $T$ periods of a rollout, is instead irreducibly sequential and is carried by a \texttt{lax.scan}.
All of these computations happen inside one \texttt{jit} boundary, enclosing the optimiser, the environment reset and every update.
The computational design just described suggests a simple model of run time, which we will later use to organise and understand our measurements.
To write it down, we first define the sequential depth $D$ as the number of irreducible stages that must run one after another. 
In practice, these are the environment periods and the gradient steps, since period $t+1$ depends on period $t$ and gradient step $j+1$ depends on step $j$.
Two hardware-related properties also need to be defined.
% %
The latency $\tau$ is the fixed overhead of launching a one-stage computation, and hence it does not depend on the width $W$, while the effective maximum throughput $P$ is the number of agent-transitions per second the device can sustain once it is fully occupied.
Given these definitions, we can model the total training time $T_{\mathrm{train}}$ as
\begin{equation}\label{eq:cost}
T_{\mathrm{train}} \;\approx\; D\left(\tau + \frac{W}{P}\right),
\end{equation}
since a training run performs $D$ stages, and each stage costs a fixed $\tau$ for launching the batch computation and then roughly $W/P$ for performing it.
The above equation clearly identifies two regimes meeting at $W \approx \tau P$, i.e. the batch width whose execution costs roughly as much as one launch overhead.
For $W \ll \tau P$ the fixed cost $D\tau$ dominates and enlarging the batch width is close to free, while for $W \gg \tau P$ the cost becomes linear in $W$ with slope $D/P$.
In practice, stages are of two kinds, rollout periods and gradient steps, each with different overheads and throughput; hence $\tau$ and $P$ are to be read as effective values, averaged over the two kinds in the fixed proportion in which they occur.

In our experiments, we control the total training budget via the number of rollout periods, with a single period consisting of $W = E \times n$ agent-transitions.

\begin{figure}
    \centering
    \vspace{-0.3cm}
    \includegraphics[width=0.85\linewidth, trim={0.0 0 0 0}, clip]{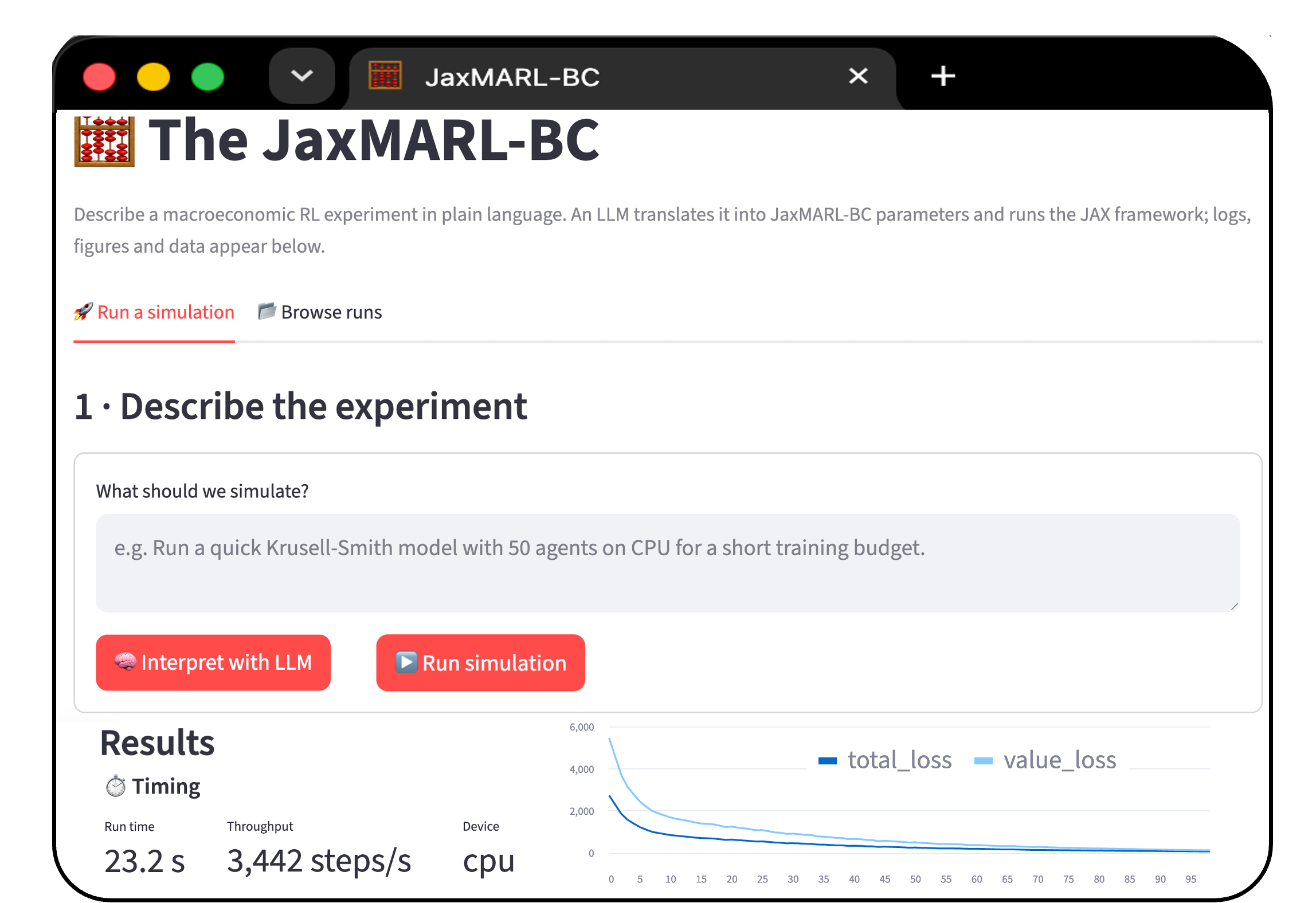}
    \vspace{-0.3cm}
    \caption{
    The LLM-augmented dashboard for \pkg{}.
    }
    \label{fig:illustration}
\end{figure}

\subsection{The LLM-augmented dashboard}
\label{subsec:dashboard}

Figure~\ref{fig:illustration} shows a screenshot of the user-friendly dashboard we built for \pkg{}.
The dashboard lets users launch a simulation from a natural-language request, so that non-technical users too can easily experiment with the simulator.
Under the hood, the natural-language request is sent to an LLM, currently either an OpenAI model via its API or a local Ollama model. 
The LLM turns the request into the YAML configuration file read by the simulation engine.
Importantly, the resulting configuration is previewed in the dashboard, and can be edited before training starts.
The dashboard then launches the training run, streams its log to the page, and renders the resulting artefacts.
We note that \pkg{} can also be run from the command line with greater parameter flexibility, and we use this approach in our experiments.
The dashboard is intended as a convenience tool for exploration, and we do not evaluate it in this work.

\section{Experimental setup}\label{sec:setup}

To validate and demonstrate the capabilities of \pkg{}, we perform three main experiments: (i) a correctness check of our simulator against the original MARL-BC results, (ii) an assessment of its scalability, and (iii) a large-scale calibration exercise that showcases its practical utility.
The general setup of all experiments is described in Sec.~\ref{subsec:general-setup}, while the detailed specifications of the three experiments are given in Sec.~\ref{subsec:three-experiments-setup}.

\subsection{General setup}
\label{subsec:general-setup}

\vspace{0.2cm}
\noindent
\textbf{Hardware and networks.}
We perform our training runs on a P100 GPU, and in experiment (ii) we also use an Intel Xeon CPU.
In all the experiments, the policy is a shared-parameter network, a $64 \times 64$ tanh MLP. 
We use PPO with the Adam optimiser, a clipping parameter of $0.2$, a GAE decay of $0.95$, a discount factor equal to $\beta$, and a learning rate of $3 \times 10^{-4}$.
Each gradient update iterates for $4$ epochs over the collected experience, split into $64$ minibatches.
%
% We prefer PPO to the SAC algorithm used in~\cite{gabriele2026}, since its on-policy updates over large batches of parallel rollouts map naturally onto the compiled design of Sec.~\ref{subsec:speedup}.
%
All computations use single-precision (\texttt{float32}) arithmetic, the JAX default on accelerators.

\vspace{0.2cm}
\noindent
\textbf{Model parameters.}
%\textbf{The Krusell--Smith calibration.}
%
\pkg{} can simulate any MARL-BC economy, and we have also checked its correctness against the $n=1$ limit of the model, which is a relatively simple limit since a single household coincides with the representative agent.
Here, however, we only report the mean-field ($n\gg1$) limit of the model, as this is much harder to replicate and it is also the limit that can most benefit from the large-scale extensions unlocked by fast training.
Specifically, we parametrise \pkg{} to reproduce the mean-field Krusell--Smith model as follows.
We set $\alpha = 0.36$, $\beta = 0.95$ and $\delta = 0.025$.
Aggregate technology follows a two-state Markov chain, $A_t = 1.02$ in good states and $A_t = 0.98$ in bad ones, with each state persisting with probability $7/8$.
Individual employment follows a second chain, where an employed household supplies $\bar{\ell} = 1.11$ units of labour and an unemployed one supplies none, with the unemployment rate at $4\%$ in good states and $10\%$ in bad ones, as in~\cite{KrusellSmith1998}.
Since the labour endowment is not chosen by the agents in the Krusell--Smith setting, we set $b = 0$ and remove labour from Eq.~\eqref{eq:reward}, and set observations to $(k^i_t, \ell^i_t, K_t, A_t, \kappa^i)$.
We append the individual capital productivity $\kappa^i$ to allow for the ex-ante heterogeneous runs of experiment (iii), while it is not needed for (i) and (ii).
Finally, we sample initial capital $k^i_0$ uniformly on~$[10, 70]$ and redraw it at every episode reset.

\subsection{The three experiments}
\label{subsec:three-experiments-setup}

\vspace{0.2cm}
\noindent
\textbf{(i) Correctness.}
We train a Krusell--Smith economy using $640$ gradient updates and using $E = 12$ parallel environments with episodes of $n = 200$ households interacting over $T = 200$ periods.
For evaluation, we set up a fresh simulation of $200$ periods with a frozen policy network, and report averages over the last $50$ periods.
The quantities we report are the ones used in~\cite{gabriele2026} in their comparison with the original Krusell--Smith model.
We then repeat the same experiment with a much larger population of $n = 100{,}000$ households.

\begin{figure*}[ht]
    \centering
    % Composite built in TikZ from the full-resolution panels (replaces the low-resolution ks-economy-combined.png)
    \begin{tikzpicture}
        \node[inner sep=0, anchor=north west] (top) at (0,0) {\includegraphics[width=0.955\linewidth]{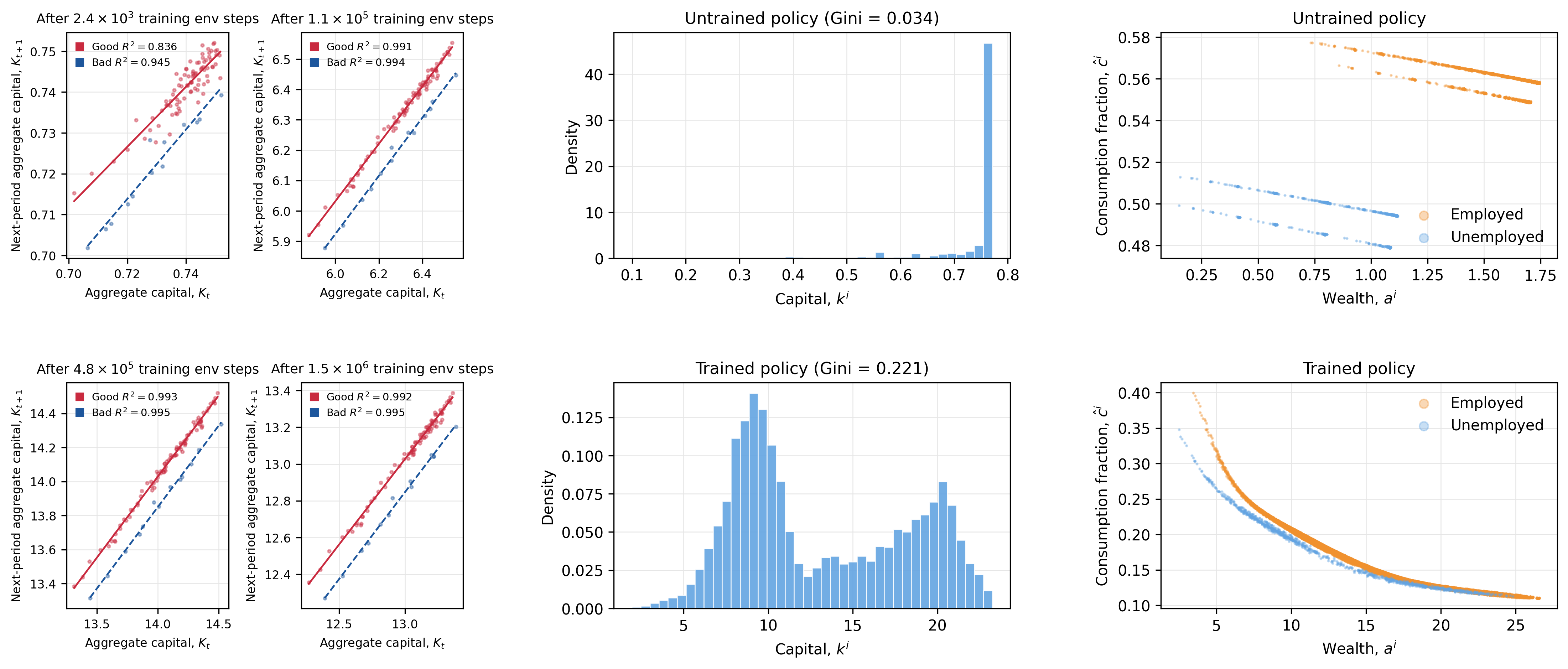}};
        \node[inner sep=0, anchor=north west] (bot) at ([yshift=-9pt]top.south west) {\includegraphics[width=0.955\linewidth]{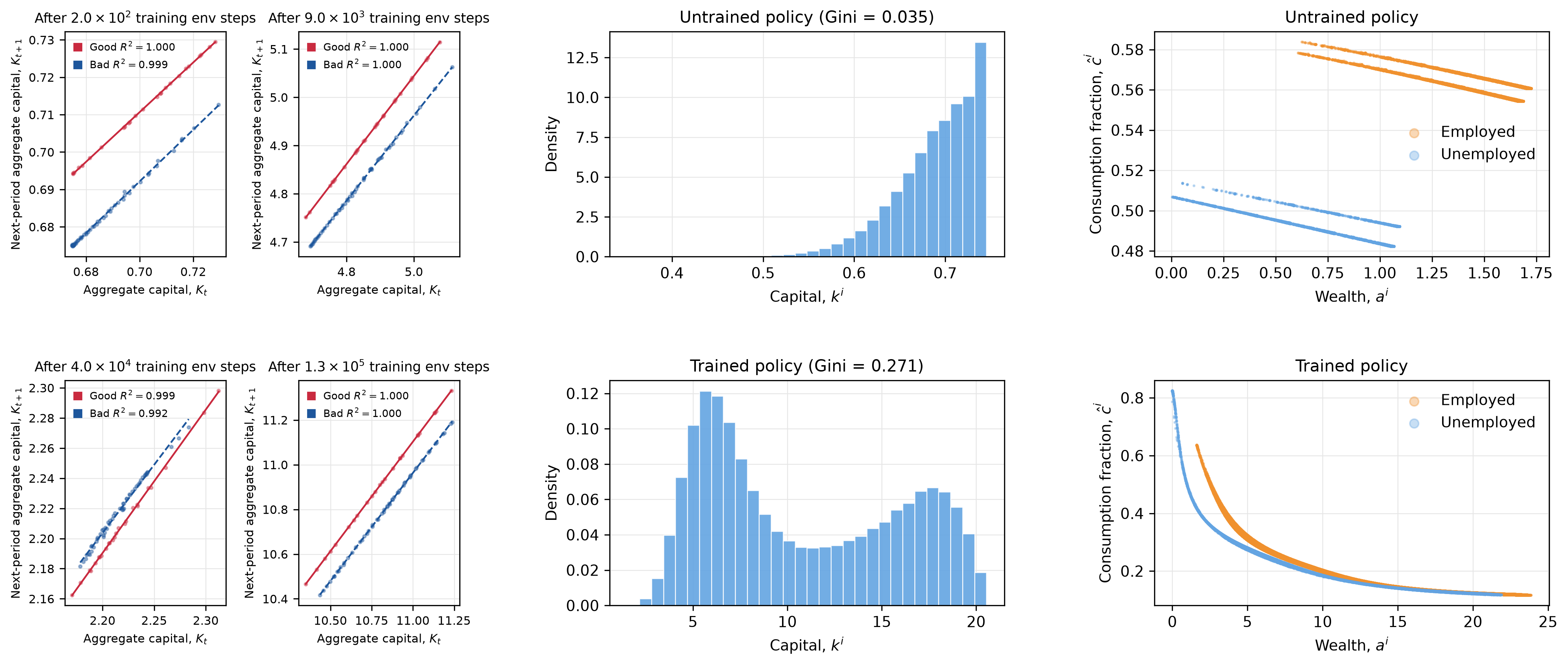}};
        \draw[line width=1.2pt, black!70] ([xshift=-0.035\linewidth, yshift=-4.5pt]top.south west) -- ([yshift=-4.5pt]top.south east);
        \draw[decorate, decoration={brace, amplitude=7pt}, thick, black!70] ([xshift=-3pt, yshift=2pt]top.south west) -- ([xshift=-3pt, yshift=-2pt]top.north west) node[midway, left=8pt, rotate=90, anchor=south, black] {$\boldsymbol{n = 200}$};
        \draw[decorate, decoration={brace, amplitude=7pt}, thick, black!70] ([xshift=-3pt, yshift=2pt]bot.south west) -- ([xshift=-3pt, yshift=-2pt]bot.north west) node[midway, left=8pt, rotate=90, anchor=south, black] {$\boldsymbol{n = 100{,}000}$};
    \end{tikzpicture}
    % \begin{subfigure}{0.99\linewidth}
    %     \centering
    %     \includegraphics[width=\linewidth]{figures/ks_economy_200.png}
    %     \caption{$n = 200$ households}
    %     \label{fig:ks-economy-200}
    % \end{subfigure}
    % \\[-1pt]
    % \begin{subfigure}{0.99\linewidth}
    %     \centering
    %     \includegraphics[width=\linewidth]{figures/ks_economy_100k.png}
    %     \caption{$n = 100{,}000$ households}
    %     \label{fig:ks-economy-100k}
    % \end{subfigure}
    \caption{Correctness check on the Krusell--Smith economy.
    \normalfont
    The Krusell--Smith economy of experiment~(i) with $n = 200$ (top) and $n = 100{,}000$ (bottom) households.
    \emph{Left}: one-step transitions of aggregate capital $(K_t, K_{t+1})$ coloured by aggregate state (good/bad), with a law of motion fitted separately in each state, at four points in training.
    \emph{Centre}: stationary distribution of individual capital, with its Gini coefficient, before and after training.
    \emph{Right}: learned consumption fractions against wealth (cash-on-hand), by employment status, before and after training.
    }
    \label{fig:ks-economy}
\end{figure*}

\begin{figure}
    \centering
    \includegraphics[width=0.99\columnwidth]{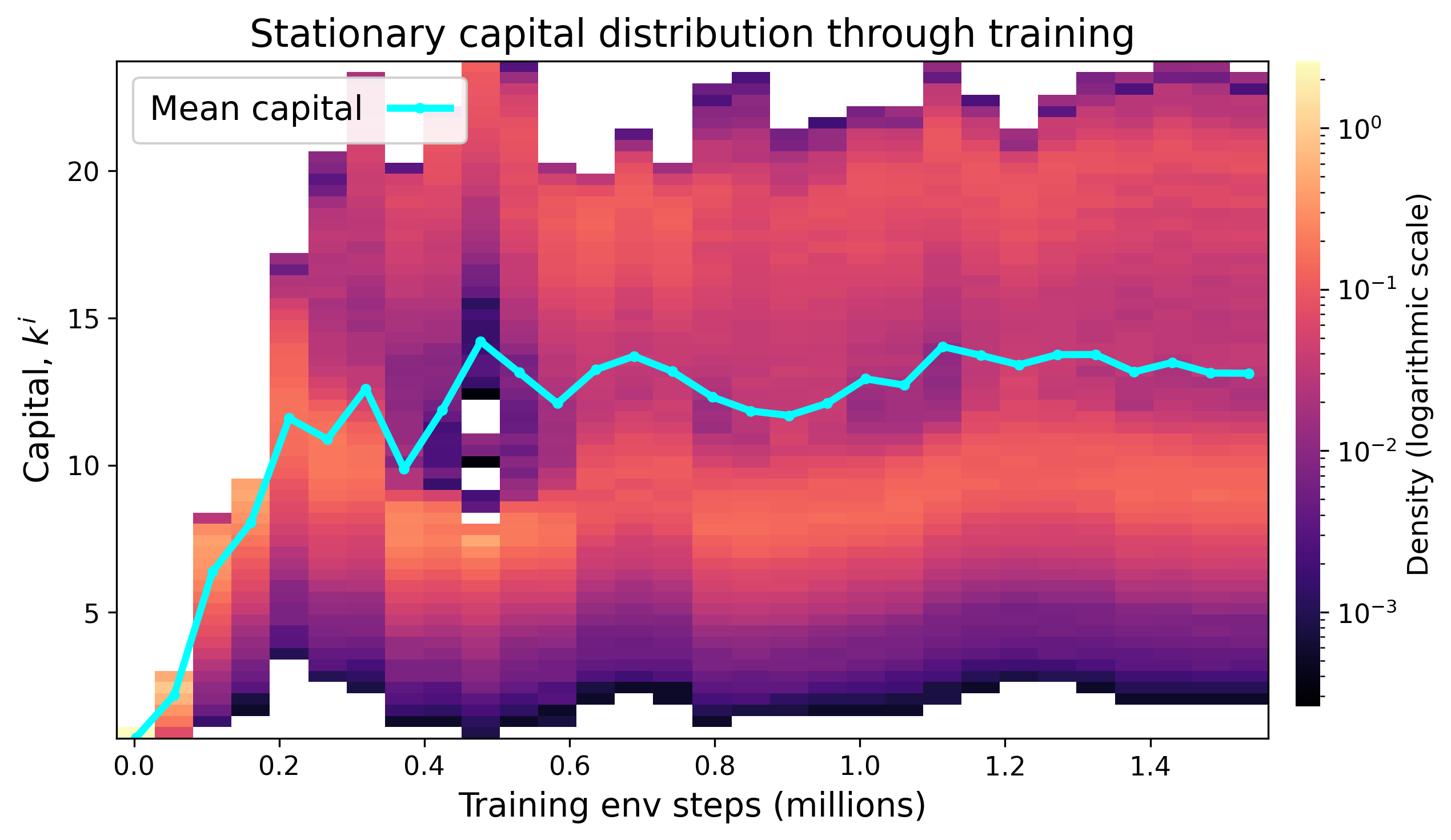}
    \caption{The capital distribution through training.
    \normalfont
    The distribution of agents' capital for $n=200$ agents, at evenly spaced training snapshots, with its mean overlaid.
    The economy quickly leaves its untrained state and settles onto the stationary distribution of Figure~\ref{fig:ks-economy}, which then holds for the rest of training.
    }
    \label{fig:wealth-vs-time}
\end{figure}

\vspace{0.2cm}
\noindent
\textbf{(ii) Scaling.}
We first compare the wall-clock training times of the original paper with those obtained by our simulator under comparable settings: a single environment ($E = 1$), a budget of $10^5$ rollout periods, and six populations, $n \in \{10, 20, 50, 100, 200, 500\}$. 
For a complete picture, we report the time of our simulator when running both on CPU and on GPU.
Then, we push our simulator to larger sizes and perform CPU and GPU training runs with up to $n=10{,}000$ and $n=200{,}000$ households, respectively.
Finally, we show how computation times depend on the batch width's dimensions $E$ and $n$.
Specifically, we let $n$ and $E$ each take the nine values $\{1, 2, 5, 10, 20, 50, 100, 200, 500\}$, giving $81$ runs, of an identical sequential depth of $10^5$ rollout periods. 
All reported times are averages over three runs.

\vspace{0.2cm}
\noindent
\textbf{(iii) Calibration of ex-ante heterogeneity.}
In the third experiment, we illustrate the practical usefulness of our simulator by performing a large-scale calibration exercise.
We train an extended model with an ex-ante different capital productivity $\kappa^i$, and demonstrate how a sweep over different levels of $\kappa^i$ heterogeneity can bring the simulation closer to the data.
Specifically, we take each $\kappa^i$ as an i.i.d.\ log-normal variable with log-scale $\sigma$ and log-location $-\sigma^2/2$ so that the population mean remains approximately $1$ for every $\sigma$.
The dispersion $\sigma$ ranges from $0$, the homogeneous benchmark $\kappa^i \equiv 1$, to $0.5$, on a grid of step $0.05$ with ten seeds at each, for a total of $110$ runs.
Each run trains $200$ households in $E = 12$ parallel environments, as in experiment (i).

\section{Results}
\label{sec:results}
The three subsections that follow correspond precisely to the three experiments described in the experimental section.

\begin{figure*}[ht!]
    \centering
    \includegraphics[width=0.85\columnwidth]{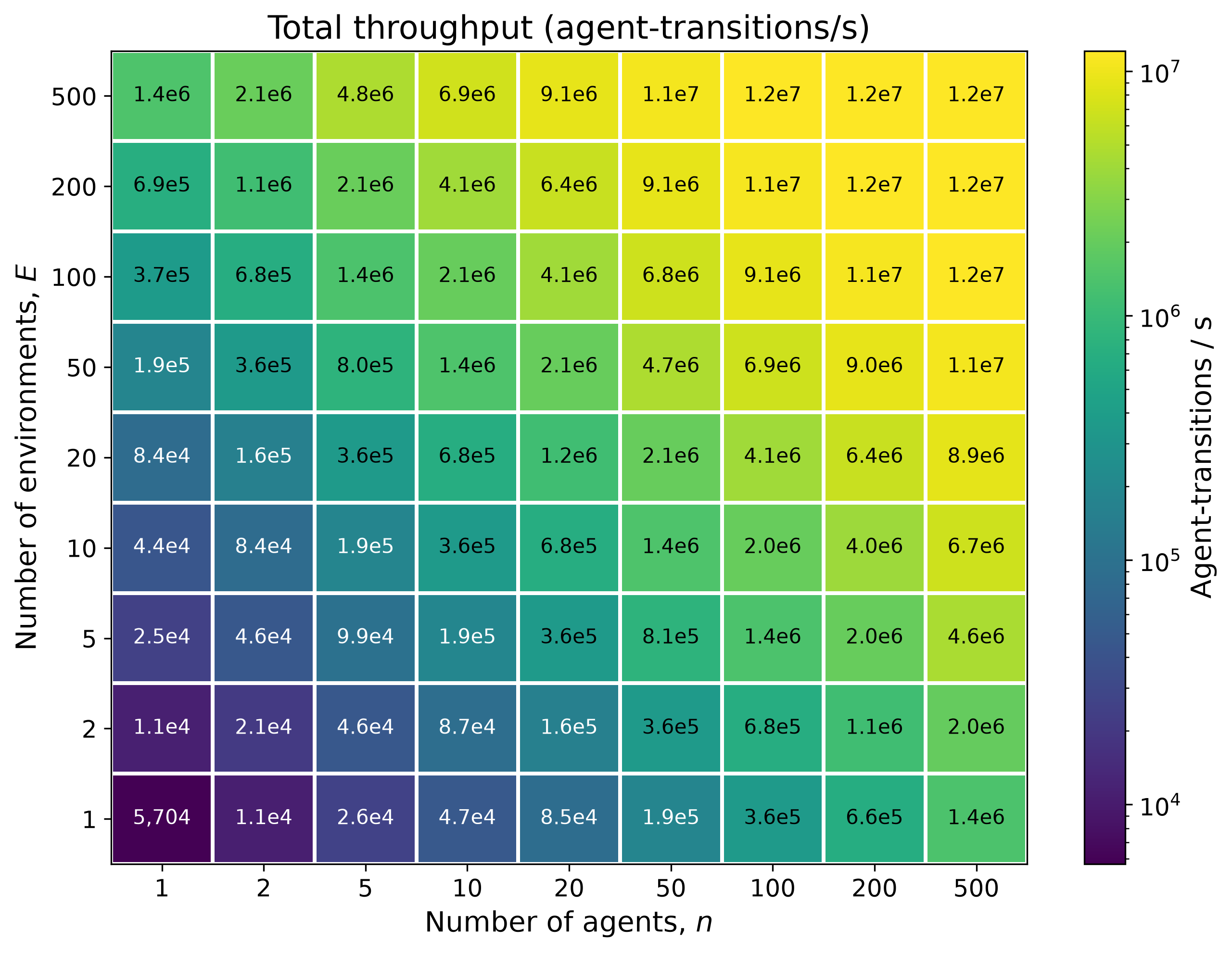}
    \hspace{0.9cm}
    \includegraphics[width=0.85\columnwidth]{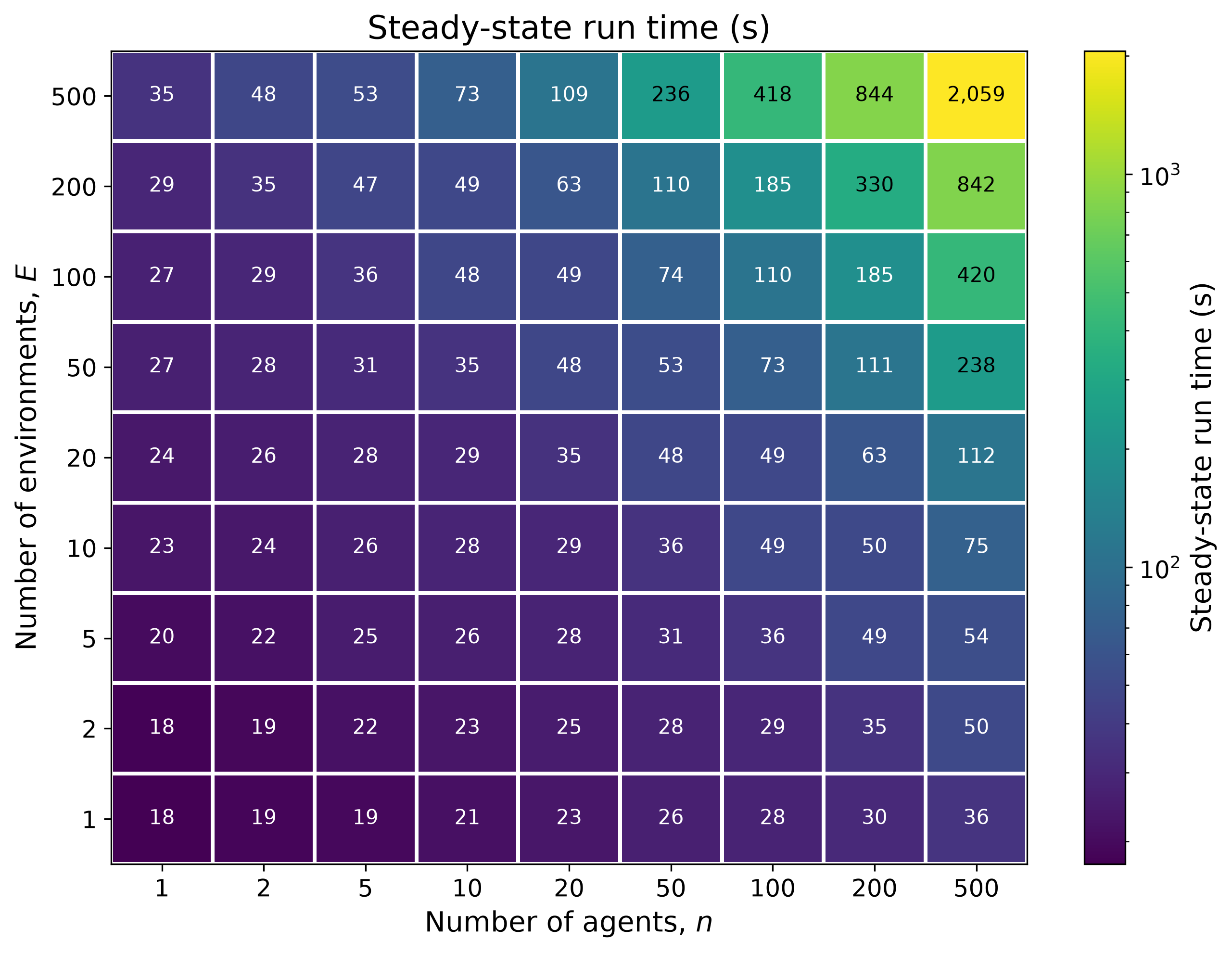}
    \caption{Throughput and run-time results.
    \normalfont
    Throughput (left) and run time (right) for a $9 \times 9$ grid over $n$ (columns) and $E$ (rows), each ranging from $1$ to $500$ at fixed sequential depth, with colour on a log scale.
    Both panels are symmetric about the diagonal, as from Eq.~\eqref{eq:cost}, throughput saturates at $\approx 10^{7}$ agent-transitions/s near $W = E \times n \approx 10^4$, past which run time grows linearly in $W$.
    }
    \label{fig:scaling-mesh}
\end{figure*}

\subsection{Correctness}
\label{sec:results-correctness}
Figure~\ref{fig:ks-economy} reports the results of \pkg{} trained to reproduce a Krusell--Smith economy with $n = 200$ (top panels) and $n = 100{,}000$ (bottom panels) households, mirroring Figure~4 of~\cite{gabriele2026}.
The figure reports the three key properties used to validate the framework: a linear law of motion for aggregate capital, a bimodal wealth distribution with a Gini index in the correct range, and a consumption fraction that declines with wealth.

Focussing on $n = 200$ first, the left panels show that the law of motion for aggregate capital becomes linear with training, the defining property of approximate aggregation in~\cite{KrusellSmith1998}.
The fit improves as the policy converges, with $R^2$ rising from $0.836$ to $0.992$ in the good state and from $0.945$ to $0.995$ in the bad state, the two states separating cleanly, and mean capital settling at $\bar K \approx 13$.
The middle and right panels show how training moves the economy from a near-degenerate untrained state, where capital collapses onto $k \approx 0.75$ and the untrained network consumes a near-constant $\approx 0.5$ fraction of wealth, to a dispersed, bimodal stationary state with a capital Gini of $0.22$ in this run.
This value is in line with the $0.20 \pm 0.05$ obtained over ten seeds in the homogeneous benchmark of Sec.~\ref{sec:results-heterogeneous-runs}.
The dispersion obtained is consistent both with~\cite{gabriele2026} and with the original Krusell--Smith calculation~\cite{KrusellSmith1998}.
The learned consumption fraction is a smooth, monotonically declining function of wealth, from $\approx 0.4$ to $\approx 0.11$, and employed households consume slightly more than unemployed households at equal wealth.

The same properties hold at $n = 100{,}000$.
With many more agents, the aggregates are averaged over a much larger population, so that all results become smoother and the law of motion for aggregate capital becomes even more clearly linear, with $R^2 = 1.000$ in both states at the end of training.
Nonetheless, the figure shows that $n = 200$ is an excellent population size for experiments, since it already captures all the phenomena of interest, while being trainable in a matter of minutes even on a CPU (see Table~\ref{tab:speedup}).

Figure~\ref{fig:wealth-vs-time} tracks the whole wealth distribution through training for the $n=200$ run.
Each of its columns is one evaluation, evenly spaced in training steps, and the colour is the density of households on a logarithmic scale.
The figure shows how the economy leaves its near-degenerate start within the first fifth of training, and then settles on a wealth distribution that remains approximately stable for the rest of the run.

\begin{table}[t]
    \centering
    \sisetup{table-format=3.0}
    \setlength{\tabcolsep}{4pt}
    \begin{tabular}{r c c c S c S}
        \toprule
        & \multicolumn{2}{c}{ours} & \multicolumn{2}{c}{PPO-CPU} & \multicolumn{2}{c}{SAC-CPU} \\
        \cmidrule(lr){2-3} \cmidrule(lr){4-5} \cmidrule(lr){6-7}
        $n$ & {GPU (s)} & {CPU (s)} & {(h)} & {($\times$)} & {(h)} & {($\times$)} \\
        \midrule
        10  & \num{21.97(3)}  & \num{13.03(28)}   & \num{0.46(14)} &  75 & \num{0.70(15)} & 115 \\
        20  & \num{23.38(10)} & \num{19.13(39)}   & \num{0.70(15)} & 108 & \num{0.77(16)} & 119 \\
        50  & \num{26.49(11)} & \num{29.89(27)}   & \num{1.49(24)} & 203 & \num{0.96(13)} & 130 \\
        100 & \num{27.98(12)} & \num{68.93(175)}  & \num{2.72(26)} & 350 & \num{1.45(9)}  & 187 \\
        200 & \num{29.63(8)}  & \num{134.14(365)} & \num{4.00(28)} & 486 & \num{1.90(22)} & 231 \\
        500 & \num{35.77(11)} & \num{276.33(279)} & \num{7.20(20)} & 724 & \num{2.99(14)} & 301 \\
        \bottomrule
    \end{tabular}
    \caption{Speedup with respect to the original implementation.
    \normalfont
    The reference times are single-CPU times of the original implementation~\cite{gabriele2026} in the same mean-field Krusell--Smith limit and at the same six population sizes.
    Our times use $E = 1$ and the same budget of $10^5$ periods, on a GPU and on a CPU backend; both are the mean over three seeds.
    Parenthesised digits give the standard deviation in units of the last digit.
    Speedups are relative to our GPU times.
    }
    \label{tab:speedup}
\end{table}

\subsection{Scaling}
\label{sec:results-scaling}

Table~\ref{tab:speedup} reports training times of \pkg{} for an increasing number of households against those of comparable training runs in~\cite{gabriele2026}.
The PPO column is the most directly comparable, since PPO is what we use in \pkg{}, and its speedup ranges from about $75\times$ at $n = 10$ to about $724\times$ at $n = 500$.
The SAC column is also relevant since SAC is the preferred algorithm in~\cite{gabriele2026}, and against it the speedup goes from $115\times$ to $301\times$.
These speedups combine the effect of the JAX with that of the GPU.
The two can be separated using our CPU times, which are between $94\times$ and $179\times$ faster, with the GPU contributing a further factor of up to $7.7\times$ at $n = 500$.
Furthermore, in contrast with the reference implementation, our GPU run time grows slowly across the table, from $22$\,s to $36$\,s for a fifty-fold larger population.

The middle and right panels of Figure~\ref{fig:scaling-vs-nagents} show throughput and wall-clock time against population size on log--log axes.
Each plot shows four curves: the original implementation under PPO (dashed red line) and under SAC (dashed orange line) on CPU, \pkg{} on CPU (dashed light blue line) and on GPU (solid blue line).
Our GPU throughput increases substantially from $\approx 10^4$ agent-transitions/s at $n \approx 1$ to $\approx 10^7$/s at $n = 10{,}000$, before plateauing and staying constant up to $n = 200{,}000$.
The CPU curve starts at higher throughput levels of $\approx 5 \times 10^4$/s but plateaus much more quickly at $\approx 2 \times 10^5$/s for $n \approx 10$.
By contrast, both reference curves stay well below $ 10^4$/s throughout.

The different throughput levels directly translate to different running times in the right panel, where both reference curves exceed one hour by $n = 100$ while \pkg{} stays of the order of tens of seconds on both GPU and CPU.
The right panel also shows that the CPU-hours the original implementation needs to train $n \approx 100$ households is enough for \pkg{} to comfortably train systems of $n \approx 200{,}000$ households, opening a regime that was previously out of reach and, with it, a whole new space of large-scale experiments with heterogeneous agents.

Figure~\ref{fig:scaling-mesh} reports again throughput and wall-clock time, but on a grid experiment with increasing population size ($n$) and number of environments ($E$).
In both panels, the colour gradient is visibly symmetric about the diagonal, consistent with the model of Eq.~\eqref{eq:cost} that expresses training cost symmetrically only via the batch width $W = E \times n$. 
The left panel shows throughput rising along the diagonal until it saturates at $\approx 10^{7}$ agent-transitions/s, a plateau reached in the neighbourhood of $W \approx 10{,}000$.
The throughput translates to computational cost in the right panel in a way that is again consistent with Eq.~\eqref{eq:cost}.
Specifically, we find that for low values of $W$ run time grows only mildly, going from $\approx 18$\,s at $W \approx 1$ to $\approx 48$\,s at $W \approx 1{,}000$, a $2.7\times$ rise across a $1{,}000\times$ rise in width, because the extra width is absorbed by throughput that is still climbing towards the plateau.
This near-constant cost can be interpreted as the latency cost $D \tau$ of Eq.~\eqref{eq:cost}.
Past that width, the device is saturated and time grows linearly with $W$, again as accounted for in Eq.~\eqref{eq:cost}, with slope $D/P$.
The two time-scaling regimes (approximately constant, and linear with $W$) are particularly visible in the right panel of Figure~\ref{fig:scaling-vs-nagents} where a linear slope is also drawn as a dashed grey line.

\begin{figure*}
    \centering
    \includegraphics[width=0.99\linewidth]{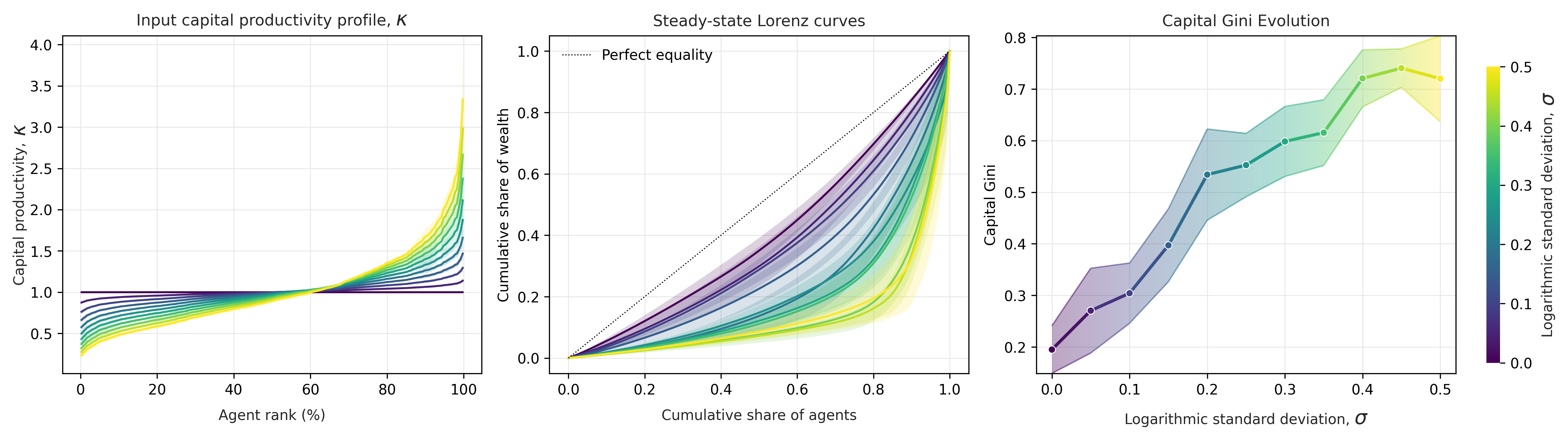}
    \caption{
    \label{fig:heterogeneous-run}
    Calibration of ex-ante heterogeneity in capital productivity.
    \normalfont
    Individual capital productivities $\kappa^i$ are drawn from a log-normal distribution with unit mean and dispersion $\sigma$, with $\sigma$ swept from $0$ (the homogeneous Krusell--Smith benchmark, $\kappa^i \equiv 1$) to $0.5$, with ten seeds per value.
    Every curve is a mean over seeds with a one-standard-deviation band.
    \emph{Left}: the realised $\kappa$ profiles by agent rank.
    Increasing $\sigma$ tilts the flat benchmark profile into an increasingly skewed one.
    \emph{Centre}: Lorenz curves of the resulting wealth distributions move away from the diagonal in the same fashion, the bottom $80\%$ of households holding $70\%$ of aggregate capital at $\sigma = 0$ and under $20\%$ at $\sigma = 0.5$.
    \emph{Right}: the resulting capital Gini, rising from $\approx 0.2$ to more than $0.7$ at $\sigma = 0.5$.
    }
\end{figure*}

\begin{figure}
    \centering
    \includegraphics[width=0.7\linewidth]{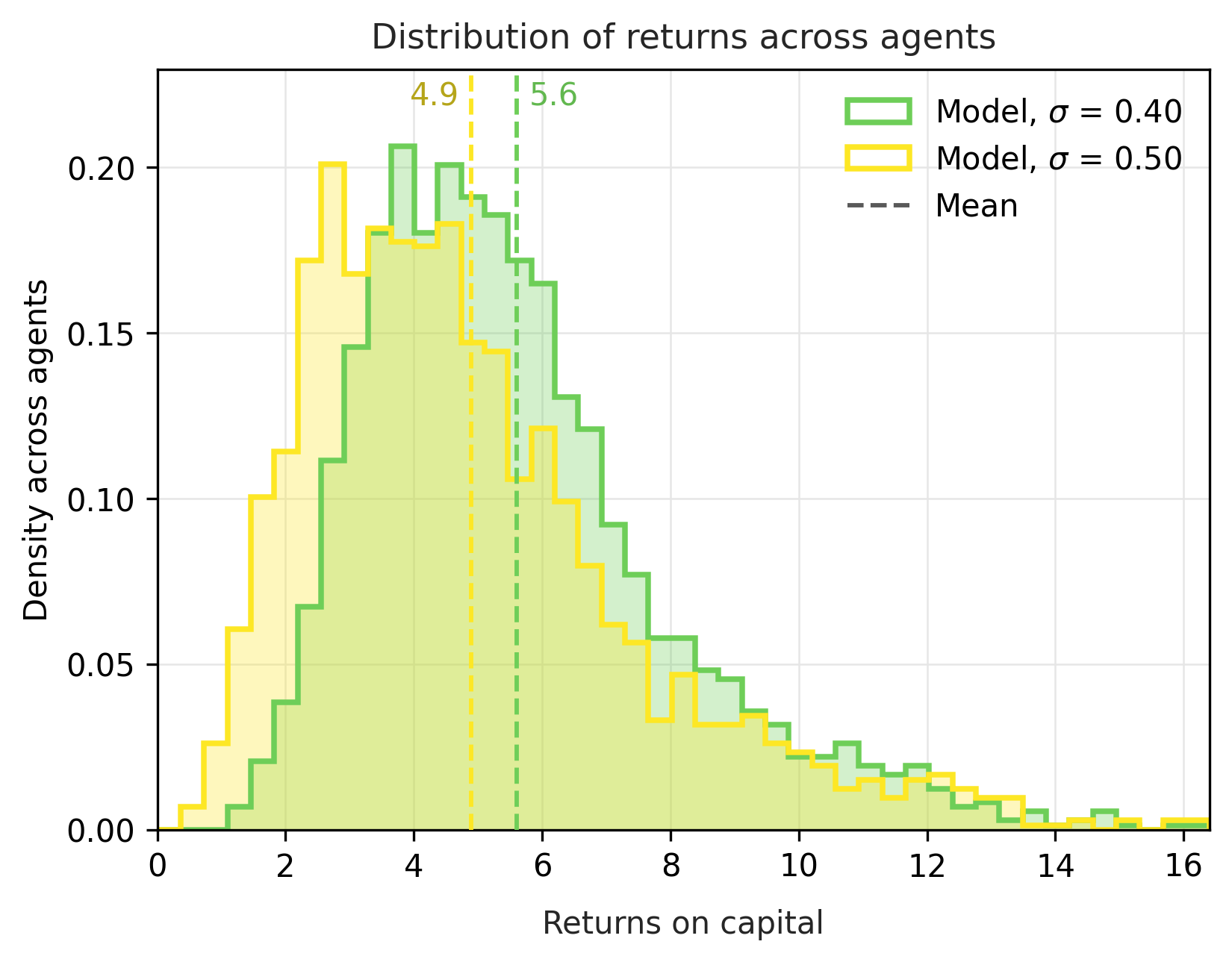}
    \caption{Heterogeneous return on capital.
    \normalfont
    Density of percentage returns on capital across agents, in percentage points, for $\sigma = 0.40$ and $\sigma = 0.50$. 
    Heterogeneity produces right-skewed distributions in which most agents earn less than the average, with a thin upper tail extending beyond returns of 15\%.
    Increasing $\sigma$ shifts mass towards lower returns and widens dispersion.
    \label{fig:heterogeneous-returns}
    }
\end{figure}

\subsection{Calibration of ex-ante heterogeneity}
\label{sec:results-heterogeneous-runs}

Having established correctness and speed, we now show how \pkg{} can be used for an advanced calibration exercise. 
Specifically, here we extend the Krusell--Smith model with heterogeneity in the ex-ante capital productivities and demonstrate how the extended model can better match empirical quantities.
We set up the model by drawing each household's capital productivity $\kappa^i$ from a log-normal distribution with dispersion parameter $\sigma$.
At $\sigma = 0$ every $\kappa^i$ equals one, and the model is equivalent to the ex-ante homogeneous Krusell--Smith economy of Sec.~\ref{sec:results-correctness}.
Since productivities enter the equation for returns on capital (see Eq.~\eqref{eq:prices}), we can calibrate their dispersion $\sigma$ to match empirically plausible
values for the Gini index and for the heterogeneous returns on capital.

Figure~\ref{fig:heterogeneous-run} reports the outcome of the training runs with means and standard deviations over seeds.
As expected, a higher dispersion $\sigma$ produces systematically more skewed Lorenz curves, and correspondingly higher Gini indices.
More specifically, the capital Gini rises from $0.20 \pm 0.05$ at $\sigma = 0$, the ex-ante homogeneous benchmark of Sec.~\ref{sec:results-correctness}, to $0.72 \pm 0.05$ at $\sigma \approx 0.4$, and it then plateaus up to $\sigma = 0.5$.
Since the wealth Gini of most advanced economies is of the order of $0.6$--$0.8$~\cite{balestra2018inequalities,saez2016wealth}, this indicates that the extended Krusell--Smith economy can be brought considerably closer to the data than the original one, and that a dispersion of $\sigma \approx 0.4$--$0.5$ constitutes a good parametrisation of the extended model.
Furthermore, Figure~\ref{fig:heterogeneous-returns} shows that the same values of $\sigma$ also imply plausible cross-sectional distributions of returns on capital, whose heterogeneity is widely recognised as an important driver of wealth concentration~\cite{fagereng2020heterogeneity,xavier2021wealth}.

This calibration exercise consists of $110$ runs, ten seeds at each of eleven dispersion values and the whole computation takes about two hours.
To give a rough estimate of the time this would have taken before \pkg{}, we can use the times in Table~\ref{tab:speedup} and assume that similar speedups also hold for the $E=12$ environments needed here.
This is a lower-bound estimate since it assumes that the reference implementation scales as efficiently as ours with $E$, and it still amounts to more than a month of compute, which shows how \pkg{} can unlock a class of calibration experiments that was previously out of reach.

\section{Conclusions}
\label{sec:conclusion}

We presented \pkg{}, a JAX-accelerated simulator for MARL-driven macroeconomic modelling. 
Training a MARL-BC economy takes tens of seconds rather than hours, and populations of $2 \times 10^5$ independent learning households come within reach of a single run, a regime never touched previously. 
Speedups of this size change what can be attempted, as large-scale runs and parameter calibration can become an integral part of model development.

Several directions follow naturally.
Wider model coverage is the most immediate extension.
The RBC and Krusell--Smith economies studied here are natural entry points, but the same machinery should accommodate Huggett economies, New Keynesian economies and their heterogeneous-agent (HANK) versions.
At the same time, the rational-expectations counterpart of the economies studied here remains within reach of classical solution methods, and it will be important to address economies with richer heterogeneity and with less stringent rational expectations or general equilibrium assumptions, where classical methods cannot be used.
On the algorithmic side, while PPO was adequate for everything we ran, richer environments may benefit from advanced off-policy learners such as SAC~\cite{haarnoja2018soft}.
A JAX simulator is also largely differentiable end-to-end, so gradients with respect to structural parameters come for free.
These could make gradient-based calibration available in place of grid search, in the spirit of recent work on differentiable agent-based models~\cite{dyer2023gradient,quera2025automatic}, and also pave the way for the development of targeted gradient-based RL solvers.

\pkg{} is released as open source, together with an LLM-augmented dashboard.
The framework is anchored to model classes the profession already trusts, and it is fast enough to quickly iterate on model specifications.
We expect these two properties to facilitate the transition of MARL from proof-of-concept demonstration to standard practice in macroeconomic modelling.

\balance
\bibliographystyle{ACM-Reference-Format}
\bibliography{bio}

\end{document}